\documentclass[twocolumn]{IEEEtran}
\usepackage{epsfig}

\def\BibTeX{{\rm B\kern-.05em{\sc i\kern-.025em b}\kern-.08em
    T\kern-.1667em\lower.7ex\hbox{E}\kern-.125emX}}

\begin{document}

\title{Current-voltage characteristics of molecular conductors: two versus three terminal}
\author{Prashant Damle, Titash Rakshit, Magnus Paulsson and Supriyo Datta 
\thanks{Corresponding author: Prashant Damle}
\thanks{Telephone: (765) 494 3383} 
\thanks{Fax: (765) 494 2706} 
\thanks{email: damle@purdue.edu}
\thanks{\copyright 2002 IEEE. Personal use of this material is permitted. However, permission to reprint/republish this material 
for advertising or promotional purposes or for creating new collective works for resale or redistribution to servers or lists, or 
to reuse any copyrighted component of this work in other works must be obtained from the IEEE.} 
\\ School of Electrical and Computer Engineering \\ Purdue University \\ West Lafayette, IN 47907}
\maketitle

\begin{abstract} 
This paper addresses the question of whether a ``rigid molecule'' (one which does not deform in an external field) used as the
conducting channel in a standard three-terminal MOSFET configuration can offer any performance advantage relative to a standard
silicon MOSFET.  A self-consistent solution of coupled quantum transport and Poisson's equations shows that even for extremely
small channel lengths (about $1~nm$), a ``well-tempered'' molecular FET demands much the same electrostatic considerations as a
``well-tempered'' conventional MOSFET. In other words, we show that just as in a conventional MOSFET, the gate oxide thickness
needs to be much smaller than the channel length (length of the molecule) for the gate control to be effective. Furthermore, we
show that a rigid molecule with metallic source and drain contacts has a temperature independent subthreshold slope much larger
than $60~mV/decade$, because the metal-induced gap states in the channel prevent it from turning off abruptly. However, this
disadvantage can be overcome by using semiconductor contacts because of their band-limited nature.
\end{abstract}
 
\begin{keywords} 
Molecular electronics, MOSFETs, electrostatic analysis, quantum transport, Non-equilibrium Green's function (NEGF) formalism.
\end{keywords}

\section{Introduction}
\label{sec:intro}

\PARstart{M}{olecules} are promising candidates as future electronic devices because of their small size, chemical tunability
and self-assembly features. Several experimental molecular devices have recently been demonstrated (for a review of the
experimental work see \cite{reed_review}).  These include two terminal devices where the conductance of a molecule coupled to
two contacts shows interesting features such as a conductance gap \cite{reed_expt}, asymmetry \cite{reichert_asymm_iv} and
switching \cite{chen}. Molecular devices where a third terminal produces a negative differential resistance \cite{lang_fet}, or
suppresses the two terminal current \cite{emberly_fet} have been theoretically studied, but most of the work on modeling the
current-voltage (IV) characteristics of molecular conductors has focused on two-terminal devices (see, for example,
\cite{datta_expt,emberly_two_pdt,diVentra,rdamle,taylor_siesta,palacios_fullerene} and references therein).

The purpose of this paper is to analyze a three-terminal molecular device assuming that the molecule behaves essentially like a
rigid solid. Unlike solids, molecules are capable of deforming in an external field and it may be possible to take advantage of
such conformational effects to design transistors with superior characteristics. However, in this paper we do not consider this
possibility and simply address the question of whether a ``rigid molecule'' used as the conducting channel in a standard
three-terminal MOSFET configuration can offer any performance advantage relative to a standard silicon MOSFET.

\begin{figure}[t!]
\centerline{\psfig{figure=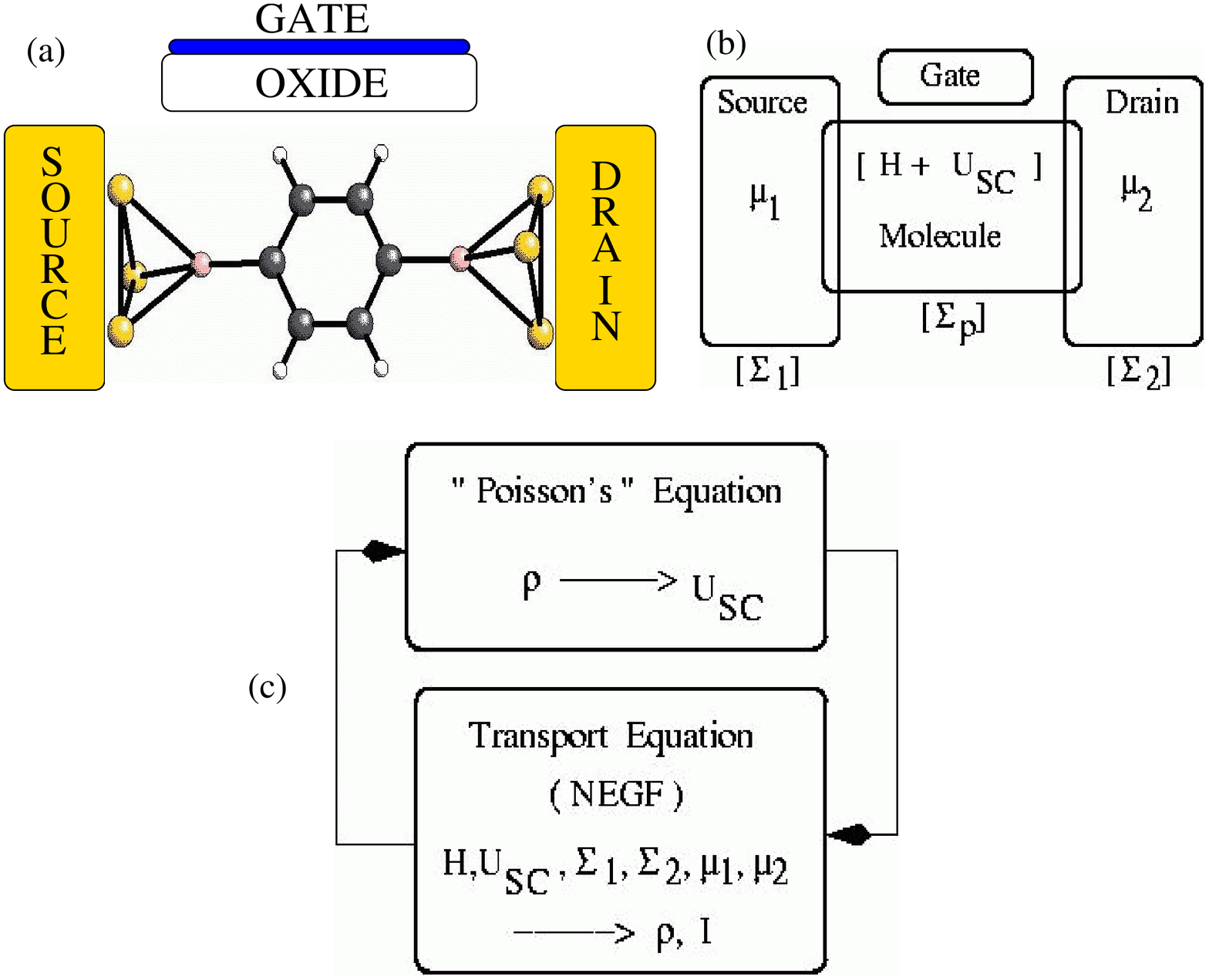,width=3in}}
\caption{
(a) Schematic of a Phenyl Dithiol molecule coupled to gold source and drain contacts. A third (gate) terminal modulates
conductance of the molecule. The phenyl ring is shown in the plane of the paper for clarity, in the actual simulation the phenyl
ring is perpendicular to the plane of the paper, ``facing'' the gate electrode. (b) The molecule is described by a Hamiltonian
$H$ and a self-consistent potential $U_{SC}$. The effect of the large contacts is described using self-energy matrices
$\Sigma_{1,2}$. Scattering processes may be described using another self-energy matrix $\Sigma_p$.  The source and drain
contacts are identified by their respective Fermi levels $\mu_1$ and $\mu_2$.  Given $H$, $U_{SC}$, $\Sigma_{1,2,p}$ and
$\mu_{1,2}$ the Non-Equilibrium Green's Function (NEGF) formalism has clear prescriptions to obtain the density matrix from
which the electron density and current may be calculated. At equilibrium (zero drain bias) $\mu_1=\mu_2=E_f$, where $E_f$ is the
common equilibrium Fermi level of the contact-molecule-contact system. When a drain bias $V_{DS}$ is applied the source and
drain Fermi levels separate by an amount equal to $qV_{DS}$ (q: electronic charge) and a current flows. This non-equilibrium
situation may be modeled by self-consistently solving the coupled NEGF-Poisson's equations as shown in (c). The word
``Poisson's'' is in quotes as a reminder that more sophisticated theories like the Hartree-Fock or the Density Functional theory
may be used to obtain $U_{SC}$.}
\label{fig:scheme}
\end{figure}

Although rigorous ab initio models are available in the literature \cite{diVentra,rdamle,taylor_siesta,palacios_fullerene}, they
normally do not account for the three-terminal electrostatics that is central to the operation of transistors. For this reason
we have used a simple model Hamiltonian whose parameters have been calibrated by comparing with ab initio models. We believe
that a simple model Hamiltonian with rigorous electrostatics is preferable to an ab initio Hamiltonain with simplified
electrostatics since the essential physics of a rigid molecular FET lies in its electrostatics.

The role of electrostatic considerations in the design of conventional silicon MOSFETs (with channel lengths ranging from
$10~nm$ and above) is well understood. For the gate to have good control of the channel conductivity, the gate insulator
thickness has to be much smaller than the channel length. Also, for a given channel length and gate insulator thickness, a
double gated structure is superior to a single gated one, simply by virtue of having two gates as opposed to one. If a molecule
is used as the channel in a standard three-terminal MOSFET configuration, the effective channel length is very small - about
$1~nm$. Would similar electrostatic considerations apply for such small channel lengths? In this paper we answer this question
in the affirmative. Specifically we will show that:
\begin{itemize}
\item{ 
The only advantage gained by using a molecular conductor for an FET channel is due to the reduced dielectric constant of the
molecular environment. To get good gate control with a single gate the gate oxide thickness needs to be less than 10\% of the
channel (molecule) length, whereas in conventional MOSFETs the gate oxide thickness needs to be less than 3\% of the channel
length \cite{taur_ning}. With a double gated structure, the respective percentages are 60\% and 20\% \cite{zhibin_ballistic}.}
\item{
Relatively poor subthreshold characteristics (a {\em temperature independent} subthreshold slope much larger than
$60~mV/decade$) are obtained even with good gate control, if metallic contacts (like gold) are used, because the metal-induced
gap states in the channel preclude it from turning off abruptly. Preliminary results with a molecule coupled to doped silicon
source and drain contacts, however, show a temperature dependent subthreshold slope ($\sim k_BT/q$). We believe this is due to
the band-limited nature of the silicon contacts, and we are currently investigating this effect.}
\end{itemize}

Overall this study suggests that superior saturation and subthreshold characteristics in a molecular FET can only arise from
novel physics beyond that included in our model. Further work on molecular transistors should try to capitalize on the
additional degrees of freedom afforded by the ``soft'' nature of molecular conductors \cite{titash} - a feature that is not
included in this study.

Although there has been no experimental report of a moleculer FET to date 
\footnote{
The authors are aware of one experimental claim (J.H. Sch\"on et al., Nature 413, page 713, 2001) reporting superior molecular
FET with a single gated geometry. This claim, however, has been strongly questioned (see article by R.F. Service in Science 298,
page 31, 2002).}
, judging from the historical development of the conventional silicon MOSFET, it is reasonable to expect that a single gated
structure would be easier to fabricate than a double gated one. With this in mind, in this paper we mainly focus on a single
gated molecular FET geometry (see Fig.~\ref{fig:scheme}). Few key results with a double gated geometry will be shown wherever
appropriate to emphasize the differences between the single and double gated structures. The paper is organized as follows:
Section~\ref{sec:theory} contains a brief description of the theoretical formulation and the simulation procedure.
Section~\ref{sec:results} presents the simulation results along with an explanation of the underlying physics.
Section~\ref{sec:conclusion} summarizes this paper.

\section{Theory}
\label{sec:theory}

A schematic figure of a molecule coupled to gold contacts (source and drain) is shown in Fig.~\ref{fig:scheme}a. As an example
we use the Phenyl Dithiol (PDT) molecule which consists of a phenyl ring with thiol (-SH) end groups. A gate terminal modulates
the conductance of the molecule.  We use a simple model Hamiltonian $H$ to describe the molecule (Fig.~\ref{fig:scheme}b). The
effect of the source and drain contacts is taken into account using self-energy functions $\Sigma_1$ and $\Sigma_2$
\cite{datta_book}.  Scattering processes may be described using another self-energy matrix $\Sigma_p$. However, in this paper we
focus on coherent or ballistic transport ($\Sigma_p=0$). The source and drain contacts are identified with their respective
Fermi levels $\mu_1$ and $\mu_2$.  Our simulation consists of iteratively solving a set of coupled equations
(Fig.~\ref{fig:scheme}c) - the Non-Equilibrium Green's Function (NEGF) formalism \cite{datta_book,datta_tut} equations for the
density matrix $\rho$ and the Poisson's equation for the self-consistent potential $U_{SC}$.  Given $H$, $U_{SC}$, $\Sigma_1$,
$\Sigma_2$, $\mu_1$ and $\mu_2$ the NEGF formalism has clear prescriptions to obtain the density matrix $\rho$ from which the
electron density and the current may be calculated. Once the electron density is calculated we solve the Poisson's equation to
obtain the self-consistent potential $U_{SC}$. The solution procedure thus consists of two iterative steps:

\begin{itemize}
\item{ {\bf Step 1}: calculate $\rho$ given $U_{SC}$ using NEGF} 
\item{ {\bf Step 2}: calculate $U_{SC}$ given $\rho$ using Poisson's equation} 
\end{itemize}	

\noindent The above two steps are repeated till neither $U_{SC}$ nor $\rho$ changes from iteration to iteration. It is worth
noting that the self-consistent potential obtained by solving Poisson's equation (Eq.~\ref{eq:poisson}) may be augmented by an
appropriate exchange-correlation potential that accounts for many electron effects using schemes like Hartree-Fock (HF) or
Density Functional Theory (DFT) \cite{szabo}. In this paper we do not consider the exchange-correlation effects.

\subsection{Step 1: To obtain $\rho$ from $U_{SC}$}

The central issue in non-equilibrium statistical mechanics is to determine the density matrix $\rho$; once it is known, all
quantities of interest (electron density, current etc.) can be calculated. A good introductory discussion of the concept of
density matrix may be found in \cite{datta_tut}. To obtain the density matrix $\rho$ from the self-consistent potential $U_{SC}$
using the NEGF formalism, we need to know the Hamiltonian $H$, the contact self-energy matrices $\Sigma_{1,2}$ and the contact
Fermi levels $\mu_{1,2}$. In this section we describe how we obtain these quantities, and then present a brief outline of the
NEGF equations.

\begin{figure}
\centerline{\psfig{figure=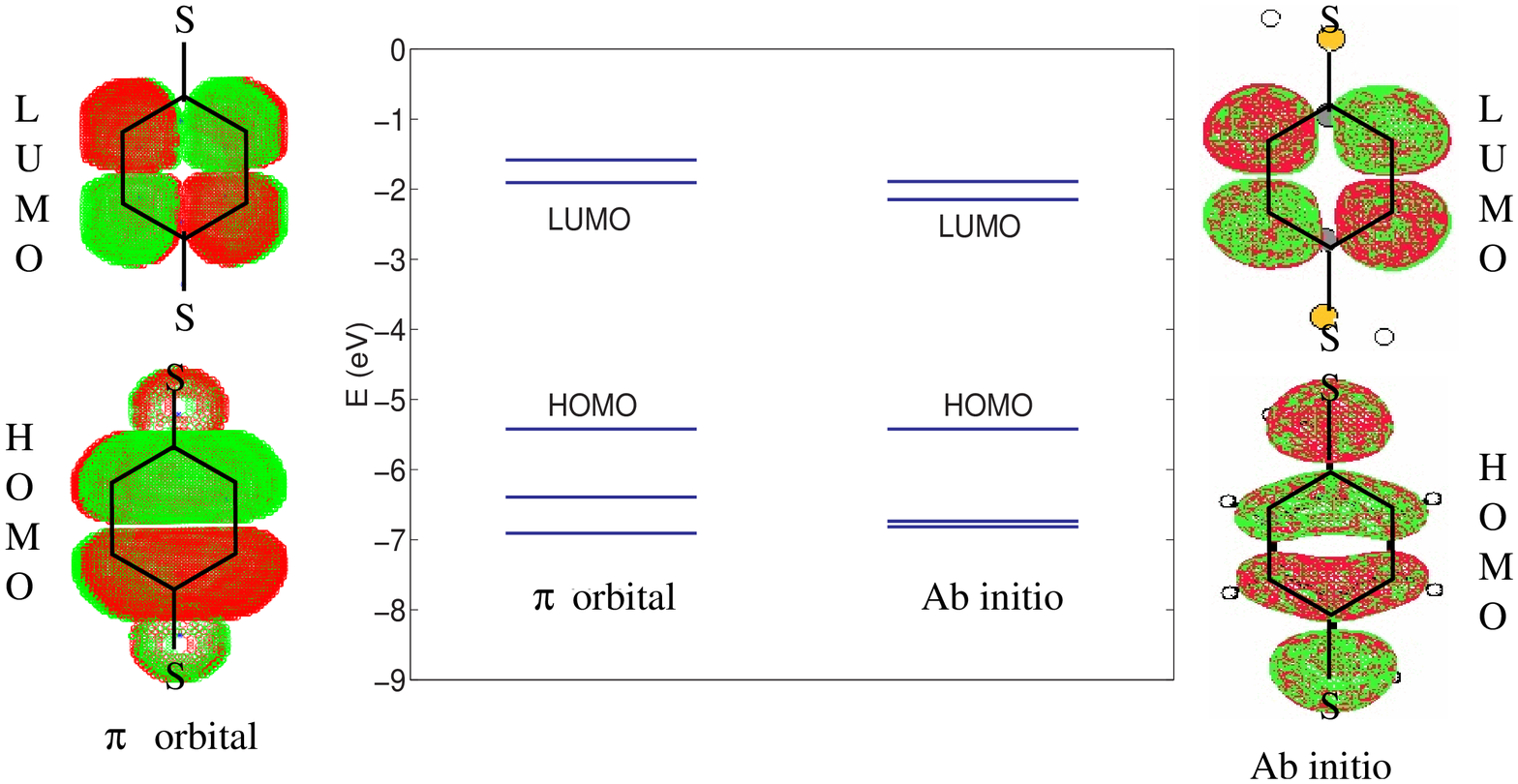,width=3in}}
\caption{
Comparison of the simple $\pi$ orbital based model with an ab initio model (density functional theory with atomic orbital basis
set)  for the PDT molecule.  The energy levels can be divided in two sets: occupied levels (analogous to the valence band) and
unoccupied levels (analogous to the conduction band). The energy gap (analogous to the bandgap) is the energy difference between
the Highest Occupied Molecular Orbital (HOMO) level (analogous to the top of the valence band) and the Lowest Unoccupied
Molecular Orbital (LUMO) level (analogous to the bottom of the conduction band). The simple $\pi$ model agrees very well with
the ab initio calculation in both the energy gap (middle) and HOMO and LUMO wavefunctions (left and right). The energy levels
obtained from the simple model were equally shifted in energy so as to make the HOMO level coincide with the ab initio HOMO
level.}
\label{fig:pi}
\end{figure}

{\em Hamiltonian}: We use a simple basis consisting of one $p_z$ (or $\pi$) orbital on each carbon and sulfur atom. It is well
known that the PDT molecule has $\pi$ conjugation - a cloud of $\pi$ electrons above and below the plane of the molecule that
dictate the transport properties of the molecule \cite{magnus_paper}.  The on-site energies of our $p_z$ orbitals correspond to
the energies of valence atomic $p_z$ orbitals of sulfur and carbon (apart from a constant shift of all levels which is allowed
as it does not affect the wavefunctions). The carbon-carbon interaction energy is $2.5~eV$ which is widely used to describe
carbon nanotubes \cite{saito_cnt_book}. The sulfur-carbon coupling of $1.5~eV$ is empirically determined to obtain a good fit to
the ab initio energy levels obtained using the commercially available quantum chemistry software Gaussian '98 \cite{gaussian}
(Fig.~\ref{fig:pi}).

Our model is very similar to the well established $p_z$ orbital based H\"uckel theory used by many quantum chemists. Although we
use a simple model Hamiltonian to describe the molecule, we believe that the essential qualitative physics and chemistry of the
molecule is captured. This is because both the molecular energy levels and the wavefunctions closely resemble those calculated
from a much more sophisticated ab initio theory (Fig.~\ref{fig:pi}).

\begin{figure}
\centerline{\psfig{figure=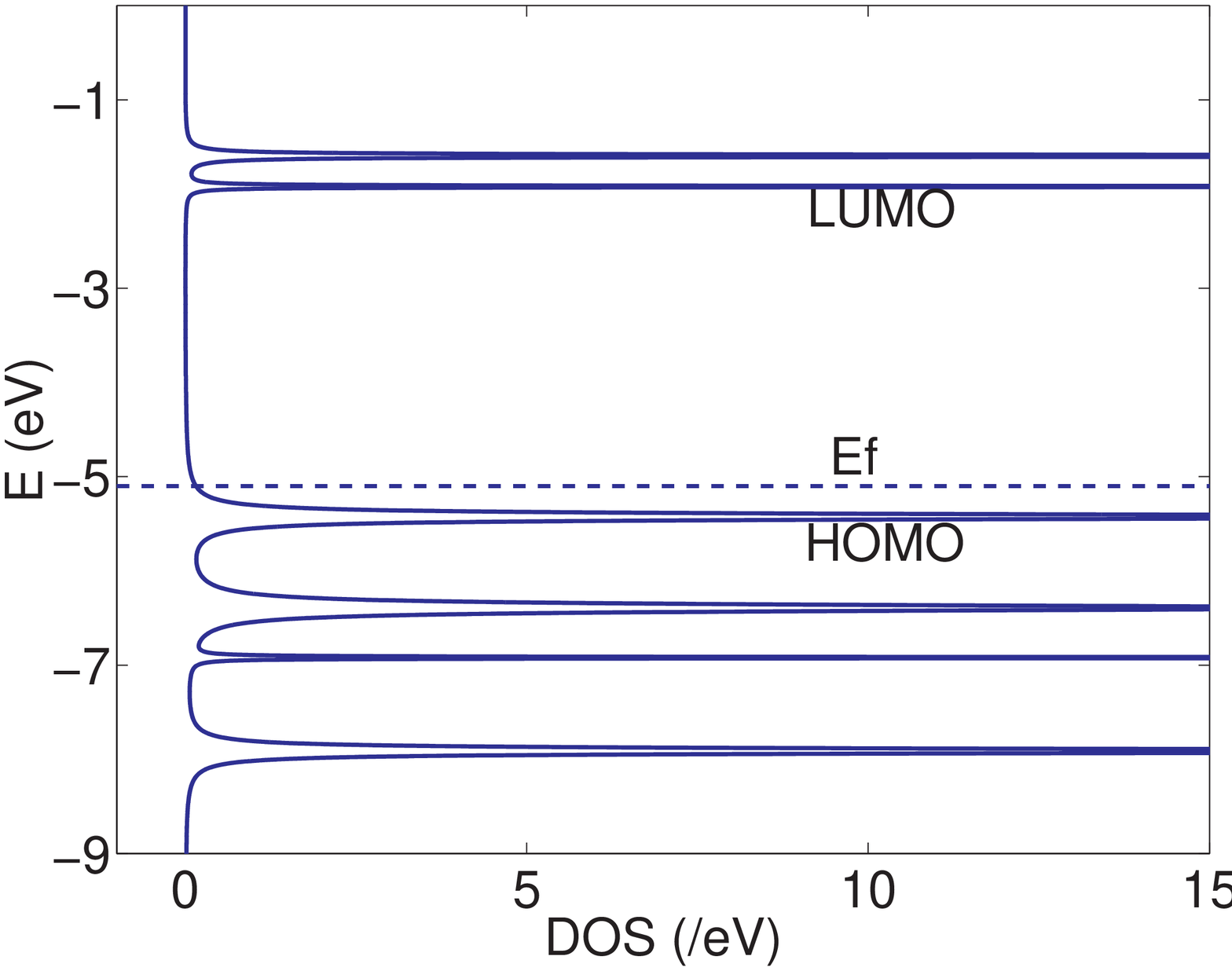,width=2.5in}}
\caption{
The discrete levels of an isolated molecule broaden into a continuous density of states (DOS) upon coupling to contacts. The
 gold FCC (111) contacts are modeled using a basis of one 's' type orbital on each gold atom. The LUMO wavefunction is
localized on the phenyl ring (Fig.~\ref{fig:pi}) and gives a sharp peak in the DOS. The HOMO is delocalized and gives a
comparatively broadened peak in the DOS. The equilibrium Fermi energy $E_f$ ($\sim -5.1~eV$ for bulk gold) lies just above the
HOMO level.}
\label{fig:dos}
\end{figure}

{\em Self-energy}: Self-energy formally arises out of partitioning the molecule-contact system into a molecular subsystem and a
contact subsystem. The contact self-energy $\Sigma$ is calculated knowing the contact surface Green's function $g$ and the
coupling between the molecule and contact $\tau$. For a molecule coupled to two contacts (source and drain) the molecular
Green's function at an energy $E$ is then written as \cite{datta_book} ($I$: identity matrix, $H$: molecular Hamiltonian,
$U_{SC}$: self-consistent potential):

\begin{equation}
G=[EI-H-U_{SC}-\Sigma_1-\Sigma_2]^{-1}
\label{eq:G}
\end{equation}
where the contact self-energy matrices are

\begin{equation}
\Sigma_{1,2}=\tau_{1,2}g_{1,2}\tau_{1,2}^\dagger
\label{eq:Sig}
\end{equation}

We model the gold FCC (111) contacts using one $s$-type orbital on each gold atom. The coupling matrix element between
neighboring $s$ orbitals is taken equal to $-4.3~eV$ - this gives correct surface density of states (DOS) of $0.07~/(eV-atom)$
for the gold (111) surface \cite{papaconst}. The site energy for each $s$ orbital is assumed to be $-8.74~eV$ in order to get
the correct gold Fermi level of $\sim -5.1~eV$.  The surface Green's function $g$ is calculated using a recursive technique
explained in detail in \cite{manoj_thesis}.  The contact-molecule coupling $\tau$ is determined by the geometry of the
contact-molecule bond.  It is believed \cite{larsen} that when a thiol-terminated molecule like Phenyl Dithiol is brought close
to a gold substrate, the sulfur bonds with three gold atoms arranged in an equilateral triangle.  For a good contact extended
H\"uckel theory predicts a coupling matrix element of about $2~eV$ between the sulfur $p_z$ orbital and the three gold $s$
orbitals. However to simulate the bad contacts typically observed in experiments \cite{reed_expt,diVentra} we reduce the
coupling by a factor of five (this factor is also treated as a parameter, and our results do not change qualitatively for a
range of values of this parameter).

Unlike the Hamiltonian, the self-energy matrices are non-Hermitian. The anti-Hermitian part of the self-energy, also known as
the broadening function:

\begin{equation}
\Gamma_{1,2}=i(\Sigma_{1,2}-\Sigma_{1,2}^\dagger)
\label{eq:Gam}
\end{equation}
is related to the lifetime of an electron in a molecular eigenstate. Thus upon coupling to contacts, the molecular
density of states (Fig.~\ref{fig:dos}) looks like a set of broadened peaks. 

{\em Where is the Fermi energy?}: When a molecule is coupled to contacts there is some charge transfer between the molecule and
the contacts, and the contact-molecule-contact system attains equilibrium with one Fermi level $E_f$. A good question to ask is
where $E_f$ lies relative to the molecular energy levels. The answer is not clear yet, the position of $E_f$ seems to depend on
what contact model one uses. A jellium model \cite{diVentra} for the contacts predicts that $E_f$ is closer to the LUMO level
for PDT whereas an extended H\"uckel theory based model \cite{tian} predicts that $E_f$ is closer to the HOMO level (see
Fig.~\ref{fig:pi} and the related caption for a description of HOMO and LUMO levels). Our ab initio model \cite{rdamle} seems to
suggest that for gold contacts, $E_f$ ($\sim -5.1~eV$) lies a few hundred millivolts above the PDT HOMO.  In this paper we will
use $E_f=-5.1~eV$ and set the molecular HOMO level (obtained from the $\pi$ model) equal to the ab initio HOMO level ($\sim
-5.4~eV$) (see Figs.~\ref{fig:pi},~\ref{fig:dos}). Once the location of the equilibrium Fermi energy $E_f$ is known, we can
obtain the source and drain Fermi levels $\mu_1$ and $\mu_2$ under non-equilibrium conditions (non-zero $V_{DS}$): $\mu_1=E_f$
and $\mu_2=E_f-qV_{DS}$.

{\em NEGF equations}: Given $H$, $\Sigma_{1,2}$, contact Fermi energies $\mu_{1,2}$ and the self-consistent potential $U_{SC}$,
NEGF has clear prescriptions \cite{datta_book} to obtain the density matrix $\rho$. The density matrix can be expressed as an
energy integral over the correlation function $-iG^<(E)$, which can be viewed as an energy-resolved density matrix:

\begin{equation}
\rho = \int dE[-iG^<(E)/2\pi] \label{rho}
\end{equation}
The correlation function is obtained from
the Green's function $G$ (eq.~\ref{eq:G}) and the broadening functions $\Gamma_{1,2}$ (eq.~\ref{eq:Gam}):
\begin{equation}
-i{G}^< = G\left({f_1\Gamma_1 + f_2\Gamma_2}\right)G^\dagger
\end{equation}
where $f_{1,2}(E)$ are the Fermi functions with electrochemical potentials $\mu_{1,2}$:
\begin{equation}
f_{1,2}(E) = \left( 1 + \exp{\left[ {{E-\mu_{1,2}}\over{k_BT}}\right]}\right)^{-1}
\end{equation} 

\newcommand{\rb}{(\vec{r})}

The density matrix so obtained can be used to calculate the electron density $n\rb$ in real space using the eigenvectors of the
Hamiltonian $\Psi_\alpha \rb$ expressed in real space:

\begin{equation}
n\rb=\sum_{\alpha,\beta} \Psi_\alpha \rb \Psi_\beta ^* \rb \rho_{\alpha \beta}
\label{eq:nofr}   
\end{equation}
The total number of electrons $N$ may be obtained from the density matrix as:

\begin{equation}
N={\rm trace}(\rho)
\label{eq:N}
\end{equation}
The density matrix may also be used to obtain the terminal current \cite{datta_book}. For coherent transport, we can simplify
the calculation of the current by using the transmission formalism where the transmission function
\cite{datta_book}:
\begin{equation}
T(E) = {\rm trace} \left[ \Gamma_1 G \Gamma_2 G^\dagger \right] 
\end{equation}
is used to calculate the terminal current
\begin{equation}
I = (2q/h)\int_{-\infty}^\infty dE~T(E)~
\left(f_1(E)-f_2(E) \right)
\end{equation}

\subsection{Step 2: To obtain $U_{SC}$ from $\rho$}

The Poisson's equation relates the real space potential distribution $U \rb$ in a system to the charge density $n \rb$. We
assume a nominal charge density $n_0 \rb$ obtained by solving the NEGF equations with $U \rb=0$ (at $V_{GS}=V_{DS}=0$). The
Poisson's equation is then solved for the {\em change} in the charge density ($n-n_0$) from the nominal value $n_0$ 
\footnote{
The potential distribution corresponding to the nominal charge density (when no drain or gate bias is applied) is included in the 
calculation of the molecular Hamiltonian \cite{datta_expt}.
} :

\begin{equation}
\vec{\nabla}\cdot\left(\epsilon\vec{\nabla}U \rb \right) = -q^2(n \rb -n_0 \rb)
\label{eq:poisson}
\end{equation}

The Poisson (or Hartree) potential $U$ may be augmented by an appropriate exchange-correlation potential $U_{xc}$. In this 
paper, we do not take into account the exchange-correlation effects ($U_{xc}=0$). We have two schemes to solve the 
Poisson's equation: 
(1) simple Capacitance Model and
(2) rigorous numerical solution over a 2D grid in real space.  

\begin{figure}
\centerline{\psfig{figure=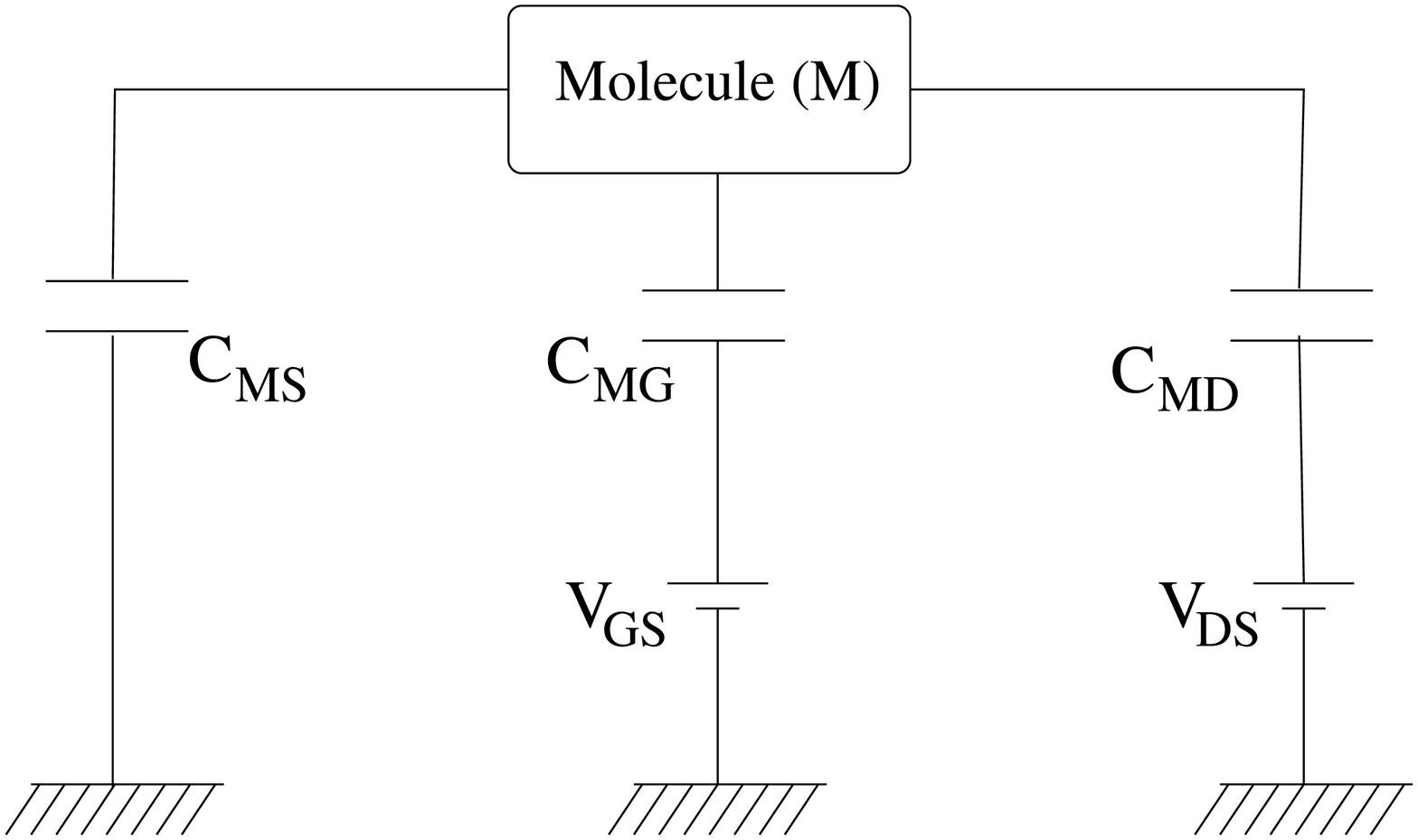,width=3in}}
\caption{Equivalent circuit model used to obtain the homogeneous (or zero charge) solution to the Poisson's equation.  The
molecular potential is controlled by the gate if the capacitative coupling $C_{MG}$ between the molecule and the gate is much
greater than the capacitative couplings $C_{MS}$ and $C_{MD}$ between the molecule and the source and drain respectively (see
Eqs.~\ref{eq:UL},~\ref{eq:beta} and related discussion).}
\label{fig:RC}
\end{figure}

{\em Capacitance Model}: We use a simplified picture of the molecule as a quantum dot with some nominal {\em total} charge $N_0$
(at $V_{GS}=V_{DS}=0$) and some average potential $U$ arising because of the {\em change} $N-N_0$ in this nominal charge due to
the applied bias. Thus $U$, $N_0$ and $N$ are numbers and not matrices. The total charge $N$ can be calculated from the NEGF
density matrix using Eq.~\ref{eq:N}. $U$ is the solution to the Poisson's equation, and may be written as the sum of two terms:
(1) A Laplace (or homogeneous) solution $U_L$ with zero charge on the molecule but with applied bias and (2) A particular (or
inhomogeneous) solution $U_P$ with zero bias but with charge present on the molecule. Thus $U=U_L+U_P$. $U_L$ is easily written
down in terms of the capacitative couplings $C_{MS}$, $C_{MD}$ and $C_{MG}$ of the molecule (Fig.~\ref{fig:RC}) with the source,
drain and gate respectively:

\begin{equation}
U_L=\beta (-qV_{GS}) + \frac{(1-\beta)}{2} (-qV_{DS})
\label{eq:UL}
\end{equation}
where 

\begin{equation}
\beta=\frac{C_{MG}}{C_{MS}+C_{MD}+C_{MG}}
\label{eq:beta}
\end{equation}
is a parameter ($0 < \beta < 1$) and is a measure of how good the gate control is. Gate control is ideal when $C_{MG}$ is 
very large as compared to $C_{MS}$ and $C_{MD}$ \footnote{
We have assumed that $C_{MS}=C_{MD}$ in eq.~\ref{eq:UL}, which is reasonable because the center of the molecule is equidistant
from the source and drain contacts in our model (see Fig.~\ref{fig:scheme}). In general, if the source (drain) is closer to the
molecule, then $C_{MS}$ ($C_{MD}$) will be bigger \cite{datta_expt}. With $C_{MS}=C_{MD}$, the molecular Laplace potential is
$V_{DS}/2$ in the absence of a gate ($\beta=0$), as is evident from eq.~\ref{eq:UL} (also see Fig.~\ref{fig:mechanism}c,d and
the related caption).
}
. In this case, $\beta=1$ and the Laplace solution $U_L=-qV_{GS}$ is essentially tied to the gate. An estimate of gate control
may be obtained from the numerical grid solution explained below by plotting $\beta$ as a function of gate oxide thickness
(Fig.~\ref{fig:beta}).

The particular solution $U_P$ may be written in terms of a charging energy $U_0$ as:

\begin{equation}
U_P=U_0(N-N_0)
\label{eq:UP}
\end{equation}
The charging energy is treated as a parameter, and may be estimated as follows. The capacitance of a sphere of radius $R$ is
$4\pi \epsilon R$. If we distribute a charge of one electron on this sphere, the potential of the sphere is $q/4\pi \epsilon R$.
For $R=1~nm$ the value of this potential is about $1.4~eV$. Thus we use a charging energy $U_0 \sim 1~eV$. $U_0$ is the charging
energy per electron per molecule and may also be estimated from the numerical grid solution by finding the average potential in
the region occupied by the molecule and carrying one electronic charge distributed equally. This numerical procedure also yields
$U_0 \sim 1~eV$ and is used to estimate the charging energy while comparing the capacitance model with the numerical grid
solution (see Fig.~\ref{fig:compare} and the related caption).

With the simple capacitance model just described, the Poisson's solution $U$ is just a number. The self-consistent potential
that adds to the $p_z$ Hamiltonian (see Eq.~\ref{eq:G}) is then calculated as $U_{SC}=UI$, where $I$ is the identity matrix of
the same size as that of the Hamiltonian.

{\em Numerical solution}: We use a 2D real space grid to solve the discretized Poisson's equation for the geometry shown in
Fig.~\ref{fig:scheme}a. The applied gate, source and drain voltages provide the boundary conditions. We use a dielectric
constant of 3.9 for silicon dioxide and 2 for the self-assembled monolayer (SAM) \cite{sam_dielectric}.

The correct procedure to obtain the real space charge density $n \rb$ (see Eq.~\ref{eq:poisson}) from the $p_z$ orbital space
density matrix $\rho$ is to make use of Eq.~\ref{eq:nofr}. However, we simplify the calculation of $n \rb$ by observing that a
carbon or sulfur $p_z$ orbital has a spread of about five to six Bohr radii (1 Bohr radius $a_B=0.529~\AA$). So for each atomic
site $\alpha$ we distribute a charge equal to $\rho_{\alpha \alpha}$ equally in a cube with side $\sim~10a_B$ centered at site
$\alpha$.

The solution to Poisson's equation yields the real space potential distribution. However, the self-consistent potential $U_{SC}$
that needs to be added to the Hamiltonian (Eq.~\ref{eq:G}) is in the $p_z$ orbital space. We assume that $U_{SC}$ is a diagonal
matrix with each diagonal entry as the value of the real space solution $U$ at the appropriate atomic position. For example, the
diagonal entry in $U_{SC}$ corresponding to the sulfur based $p_z$ orbital would be equal to $U(\vec{r}_S)$ where $\vec{r}_S$ is
the position vector of the sulfur atom.

\section{Results}
\label{sec:results}

The self-consistent procedure (Fig.~\ref{fig:scheme}c) is done with the two types of Poisson solutions discussed above. The
simple capacitance model is fast while the 2D numerical solution is slow but more accurate. The capacitance model has two
parameters, namely $\beta$ which is a measure of the gate control, and $U_0$ which is the charging energy.  These parameters can
be extracted using the 2D numerical solution. We will first present results with the capacitance model by assuming ideal gate
control, or $\beta=1$. This ideal case is useful to explain the current saturation mechanism.  We will then compare the results
obtained from the capacitance model with those obtained from the numerical solution, and show that the two match reasonably
well.

\subsection{Ideal gate control, on state} 

\begin{figure}
\centerline{\psfig{figure=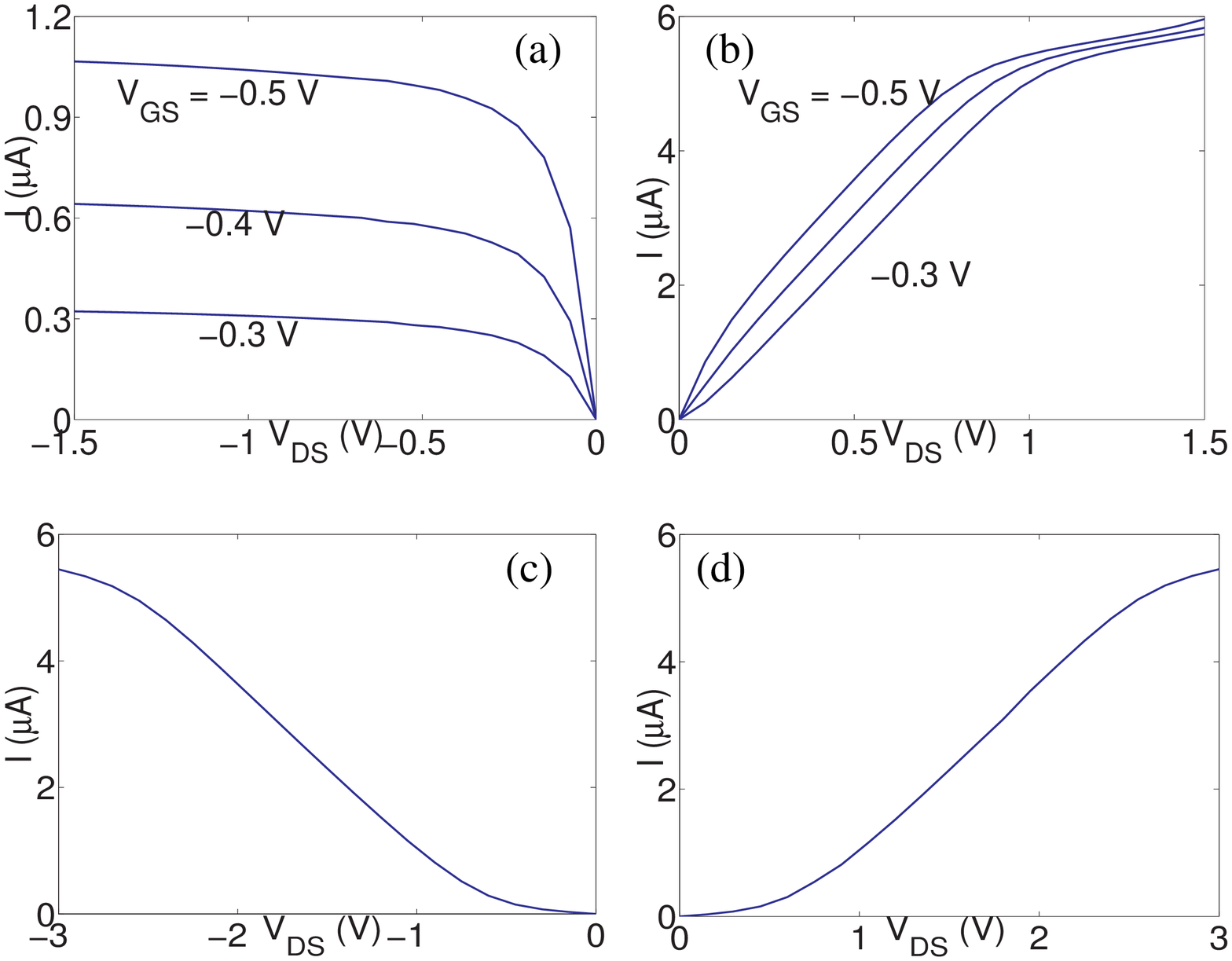,width=3in}}
\caption{Three terminal molecular drain current vs. drain to source voltage characteristic with (a) Ideal gate control
($\beta=1$), negative drain bias, (b) Ideal gate control, positive drain bias, (c) No gate control ($\beta=0$), negative drain
bias and (d) no gate control, positive drain bias. With ideal gate control the IV is asymmetric with respect to drain bias. Good
saturation and gate modulation is seen for negative drain bias but not for positive drain bias. With no gate control the IV is
symmetric with respect to drain bias. For an explanation of the underlying mechanism for each of these IV curves, see
Fig.~\ref{fig:mechanism}.}
\label{fig:ideal_iv}
\end{figure}

Fig.~\ref{fig:ideal_iv} shows the molecular IV characteristic obtained by self-consistently solving the coupled NEGF -
capacitance model Poisson's equations. We contrast the IV for ideal gate control ($\beta=1$, Fig.~\ref{fig:ideal_iv}a,b) with
that for no gate control ($\beta=0$, Fig.~\ref{fig:ideal_iv}c,d). For each case, we have shown the IV for positive as well as
negative drain voltage. We observe the following:

\begin{itemize}
\item{With ideal gate control the IV is asymmetric with respect to the drain bias. For positive drain bias, we see very 
little gate modulation of the current. For negative drain bias we see current saturation and good gate modulation - the IV 
looks like that of a MOSFET.}
\item{With no gate control the IV is symmetric with respect to the drain bias. There is no gate modulation.}
\end{itemize}

\begin{figure}
\centerline{\psfig{figure=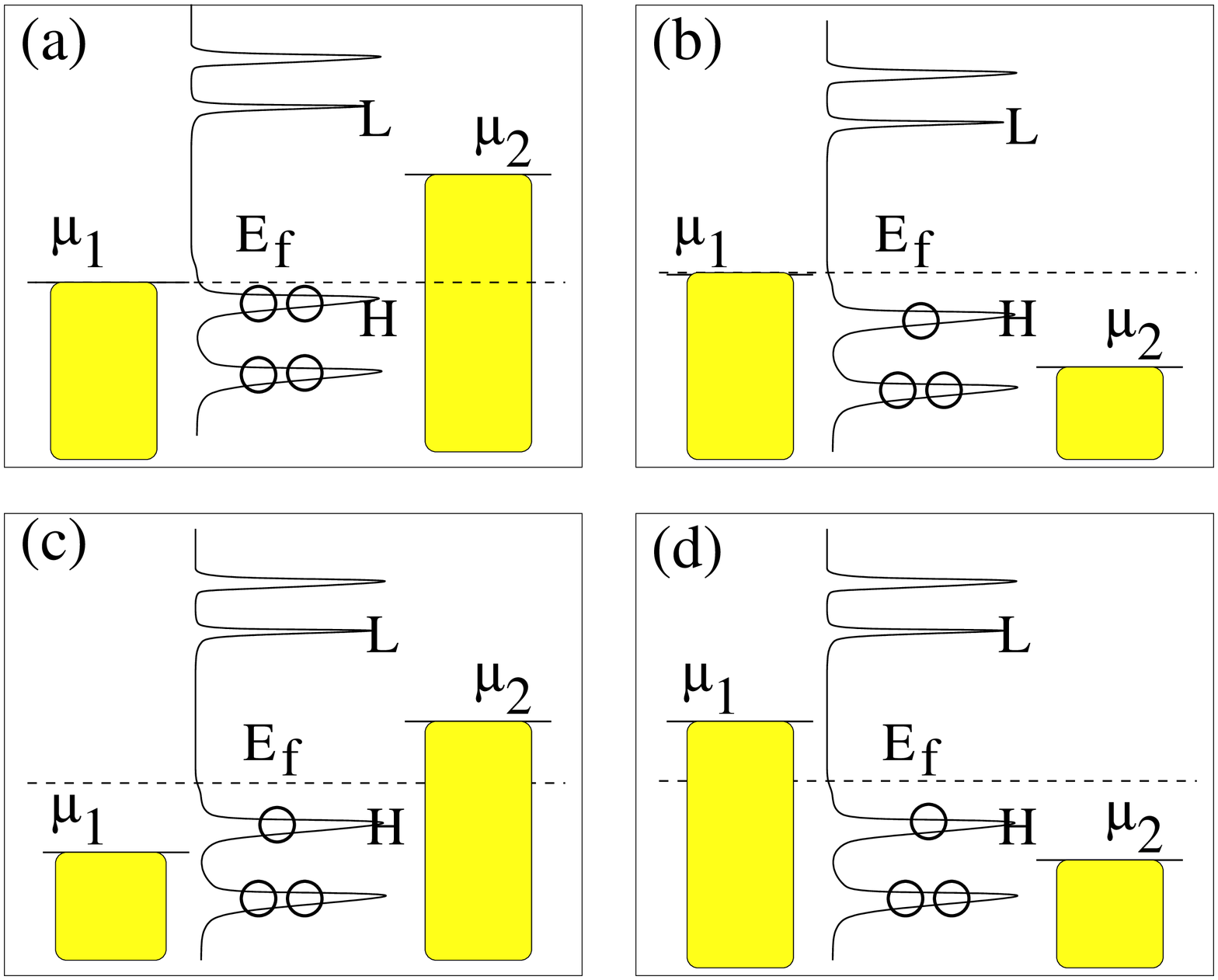,width=3in}}
\caption{
Gate induced current saturation mechanism: Assuming that the gate is very close to the molecule (ideal gate control, $\beta=1$),
the gate holds the molecular DOS fixed relative to the source Fermi level $\mu_1$ because the gate is held at a fixed potential
with respect to the source. When a negative drain bias is applied (top left), the drain Fermi level $\mu_2$ moves up relative to
the molecular DOS. Since the DOS dies out in the gap, for sufficiently high drain bias, no more DOS comes in the $\mu_1$-$\mu_2$
window and the current saturates. When a positive drain bias is applied (top right), $\mu_2$ moves down relative to DOS and
eventually crosses the HOMO. The IV is thus asymmetric. If the gate is far away (no gate control, $\beta=0$), the DOS lies
roughly halfway between the source and drain Fermi levels. In this case, for negative drain bias (bottom left), $\mu_1$ crosses
HOMO while for positive drain bias (bottom right) $\mu_2$ crosses HOMO. No gate modulation is seen as expected, and the current
is symmetric with respect to drain bias.}
\label{fig:mechanism}
\end{figure}

These features of the IV characteristic may be understood as follows (Fig.~\ref{fig:mechanism}). Let us first consider the ideal
gate case. Since the gate is held at a fixed potential {\em with respect to the source}, the molecular DOS does not shift
relative to the source Fermi level $\mu_1$ as the drain bias is changed \footnote{
This is true as long as the charging energy $U_0 \sim 1~eV$, which is typically the case. For high charging energies the
particular solution $U_P$ can dominate the Laplace solution $U_L$ (see eqs.~\ref{eq:UL},\ref{eq:UP} and related discussion),
thereby reducing gate control.
}
. For negative drain bias (Fig.~\ref{fig:mechanism}a), the drain Fermi level $\mu_2$ moves up (towards the LUMO) with respect to
the molecular DOS. Since the drain current depends on the DOS lying between the source and drain Fermi levels, the current
saturates for increasing negative drain bias because the tail of the DOS dies out as the drain Fermi level moves towards the
LUMO. If the gate bias is now made more negative, the molecular levels shift up relative to the source Fermi level, thereby
bringing in more DOS in the energy range between $\mu_1$ and $\mu_2$ (referred to as the $\mu_1$-$\mu_2$ window from now on) ,
and the current increases. Thus we get current saturation and gate modulation.

For positive drain bias (Fig.~\ref{fig:mechanism}b), $\mu_2$ moves down (towards the HOMO) with respect to the molecular DOS.
The current increases with positive drain bias because more and more DOS is coming inside the $\mu_1$-$\mu_2$ window as $\mu_2$
moves towards the HOMO peak. Once $\mu_2$ crosses the HOMO peak, the current levels off. This is the resonant tunneling
mechanism. If the gate bias is now made more negative, no appreciable change is made in the DOS inside the $\mu_1$-$\mu_2$
window, and the maximum current remains almost independent of the gate bias.

Now let us contrast this with the case where no gate is present. In this case, due to the applied drain bias $V_{DS}$, the
molecular DOS floats up by roughly $-qV_{DS}/2$ with respect to the source Fermi level. For either negative
(Fig.~\ref{fig:mechanism}c) or positive (Fig.~\ref{fig:mechanism}d)  drain bias, the current flow mechanism is resonant
tunneling. Since the equilibrium Fermi energy lies closer to the HOMO, for negative drain bias $\mu_1$ crosses HOMO while for
positive drain bias $\mu_2$ crosses HOMO \cite{datta_expt,toymodel}. No gate modulation is seen as expected, and the IV is
symmetric with respect to $V_{DS}$.

\subsection{Ideal gate control - off state} 

\begin{figure}
\centerline{\psfig{figure=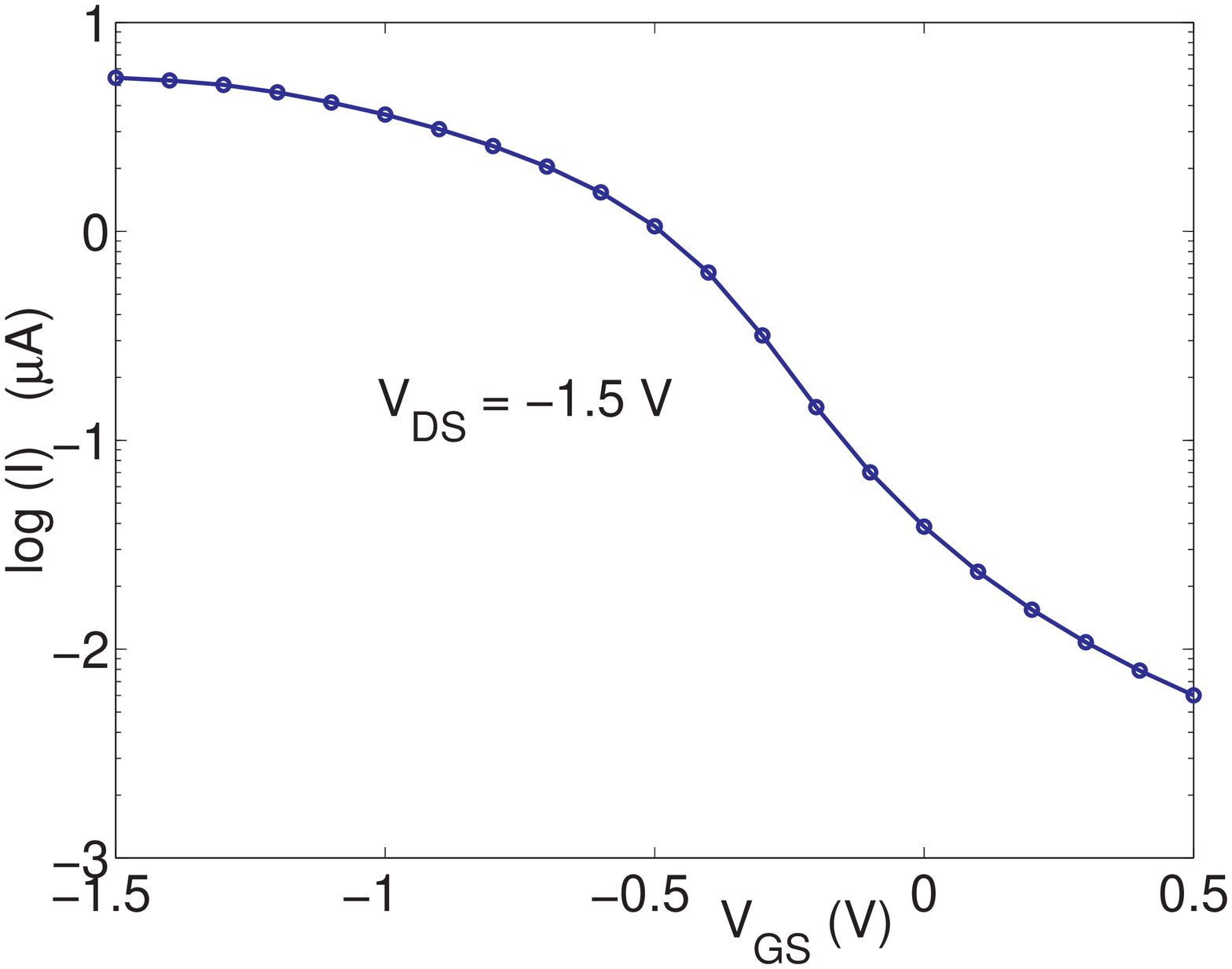,width=3in}}
\caption{
Subthreshold IV characteristic assuming ideal gate control. The {\em temperature independent} subthreshold slope is
$\sim~300~mV/decade$ even with an ideal gate. This is because the DOS in the HOMO-LUMO gap dies out very slowly as a function of
energy. This slow fall of the DOS may be attributed to the gold metal-induced gap states. Thus a molecular FET with a rigid
molecule acting as the channel is a very poor switch.}
\label{fig:subth}
\end{figure}

Fig.~\ref{fig:subth} shows the log scale drain current as a function of gate bias at high drain bias. We note that despite
assuming ideal gate control, the subthreshold slope of this molecular FET is about $300~mV/decade$ which is much worse than the
ideal room temperature $k_BT/q=60~mV/decade$ that a good MOSFET can come close to achieving. It is also worth noting here that
our simulation is done at low temperature - the subthreshold slope of the molecular FET is {\em temperature independent} and
only depends on the molecular DOS as explained below.

The poor subthreshold slope may be understood as follows. As the gate voltage is made more positive, the molecular DOS shifts
down with respect to the $\mu_1$-$\mu_2$ window. The HOMO peak thus moves away from the $\mu_1$-$\mu_2$ window, and fewer states
are available to carry the current. The rate at which the current decreases with increasing positive gate bias thus depends on
the rate at which the tail of the DOS in the HOMO-LUMO gap dies away with increasing energy (Fig.~\ref{fig:dos}).  Typically we
find that the tail of the DOS dies away at the rate of several hundred milli electron-volts of energy per decade, and this slow
fall in the DOS determines the subthreshold slope.  The slow fall in the molecular DOS may be attributed to the metal-induced
gap (MIG) states - the gold source and drain contacts have a sizeable DOS near the Fermi energy, and are separated only by a few
angstroms \footnote{
For ballistic silicon MOSFETs, due to the band-limited nature of the doped silicon source/drain contacts, the MIG DOS is
negligible. The subthreshold slope at a finite temperature is thus determined by the rate at which the difference in the source
and drain {\em Fermi function tails} falls as a function of energy. This rate depends on the temperature, and the subthreshold
slope is thus proportional to $k_BT/q$ ($\approx 60~mV$ at room temperature) for ballistic Si MOSFETs \cite{zhibin_ballistic}.
Preliminary results for a molecular FET with doped silicon source and drain contacts do show a subthreshold slope proportional
to $k_BT/q$; we are currently investigating this effect.
}. 
Since the molecule is assumed to be rigid, the molecular DOS has no temperature dependence and hence neither does the
subthreshold slope.  Thus the molecular FET with a rigid molecule acting as the channel is a very poor switching device even
with ideal gate control.

\subsection{Estimate of Gate Control}

\begin{figure}
\centerline{\psfig{figure=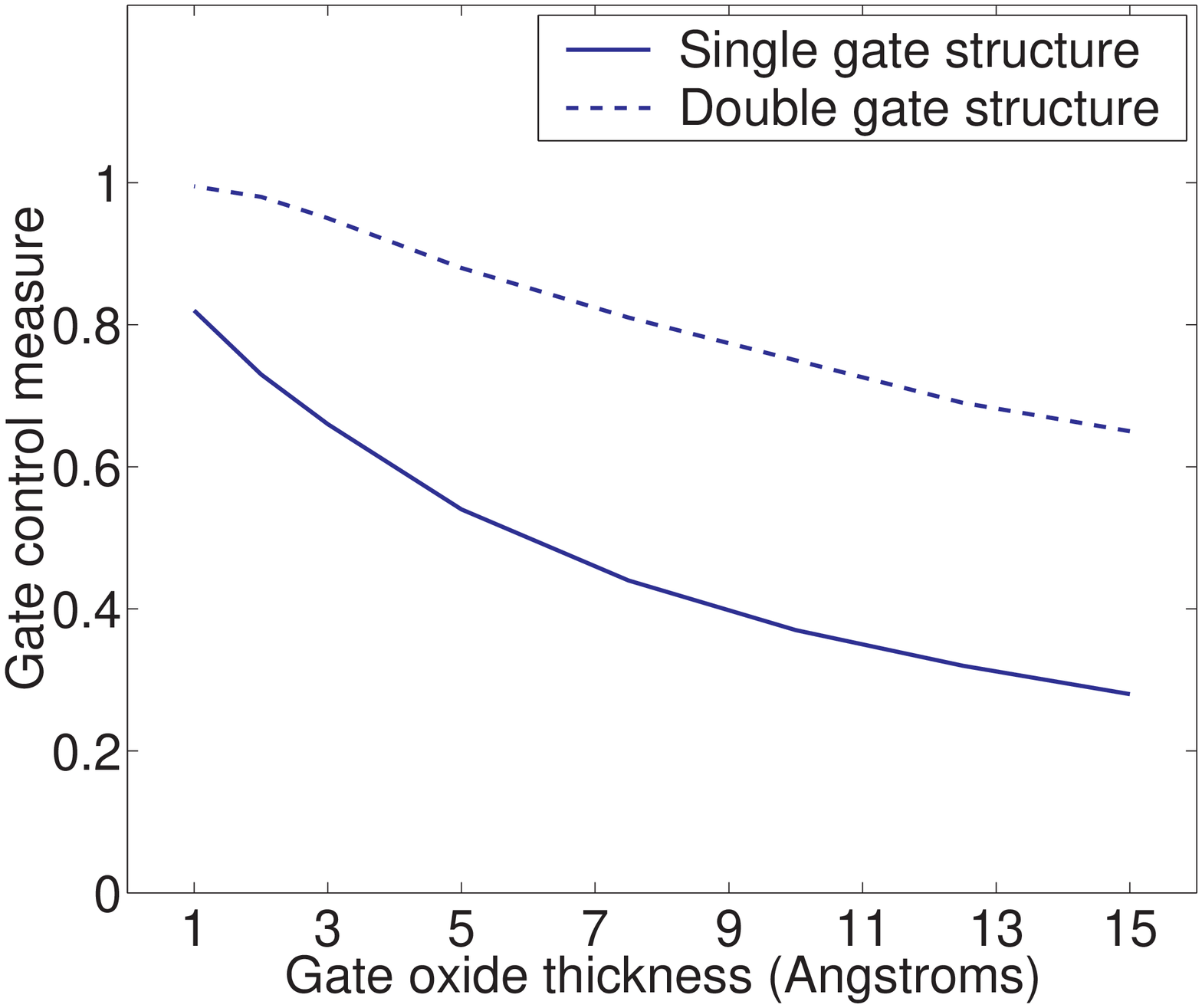,width=3in}}
\caption{
Estimate of gate control using 2D numerical Laplace's solution with a single gated geometry (solid line) and a double gated
geometry (dashed line). $\beta$ (which is a measure of gate control, see Eqs.~\ref{eq:UL},~\ref{eq:beta},~\ref{eq:beta_num} and
related discussion) is plotted as a function of the gate oxide thickness $T_{ox}$. The length of PDT molecule (equal to the
channel length of the molecular FET) is about $1~nm$. Thus in order to get good gate control ($\beta > 0.8$) the gate oxide
thickness has to be about one tenth of the channel length, or about $1~\AA$! We have used 3.9 and 2 as the dielectric constants
for silicon dioxide and the self-assembled monolayer (SAM) respectively. For a double gated geometry, good gate control can be
obtained with more realistic oxide thicknesses ($\sim ~ 7~\AA$), as expected simply because two gates can better control the
channel than one.}
\label{fig:beta}
\end{figure}

The 2D numerical Poisson's solution may be used to estimate the gate control as follows.  From Eq.~\ref{eq:UL} we see that

\begin{equation}
\beta=\left. -\frac{1}{q} ~ \frac{\partial U_L}{\partial V_{GS}}\right | _{V_{DS}}
\label{eq:beta_num}
\end{equation}
Thus $\beta$ may be estimated from the numerical solution by slightly changing $V_{GS}$ (keeping $V_{DS}$ constant) and
calculating how much the Laplace's solution changes over the region occupied by the molecule. A plot of $\beta$ calculated using
this method as a function of the gate oxide thickness $T_{ox}$ is shown in Fig.~\ref{fig:beta}.

Knowing that the channel length (length of the PDT molecule) is about $1~nm$, It is evident from Fig.~\ref{fig:beta} that in
order to get good gate control ($\beta > 0.8$) the gate oxide thickness ($T_{ox}$) needs to be about one tenth of the channel
length ($L_{ch}$), or about $1~\AA$! Thus we need $L_{ch}/T_{ox} \sim 10$ to get a good molecular FET. In a well-designed
conventional bulk MOSFET, $L_{ch}/T_{ox} \sim 40$ \cite{taur_ning}. This difference between a molecular FET and a conventional
FET may be understood by noting that the dielectric constant of the molecular environment (=2) is about 6 times smaller than
that of silicon (=11.7) \cite{lundstrom_private}.

Fig.~\ref{fig:beta} also shows $\beta$ as a function of $T_{ox}$ calculated using the 2D numerical Laplace's solution over a
double gated molecular FET structure. In this case, we find that to get good gate control, we need $L_{ch}/T_{ox} \sim 1.6$.  
Thus a double gated structure is superior to a single gated one for a given $L_{ch}$ and $T_{ox}$, as is also expected for
conventional silicon MOSFETs. The reason for this is simply that two gates can better control the channel than one.

\subsection{Comparison: Capacitance model vs. Numerical Poisson's solution} 

\begin{figure}
\centerline{\psfig{figure=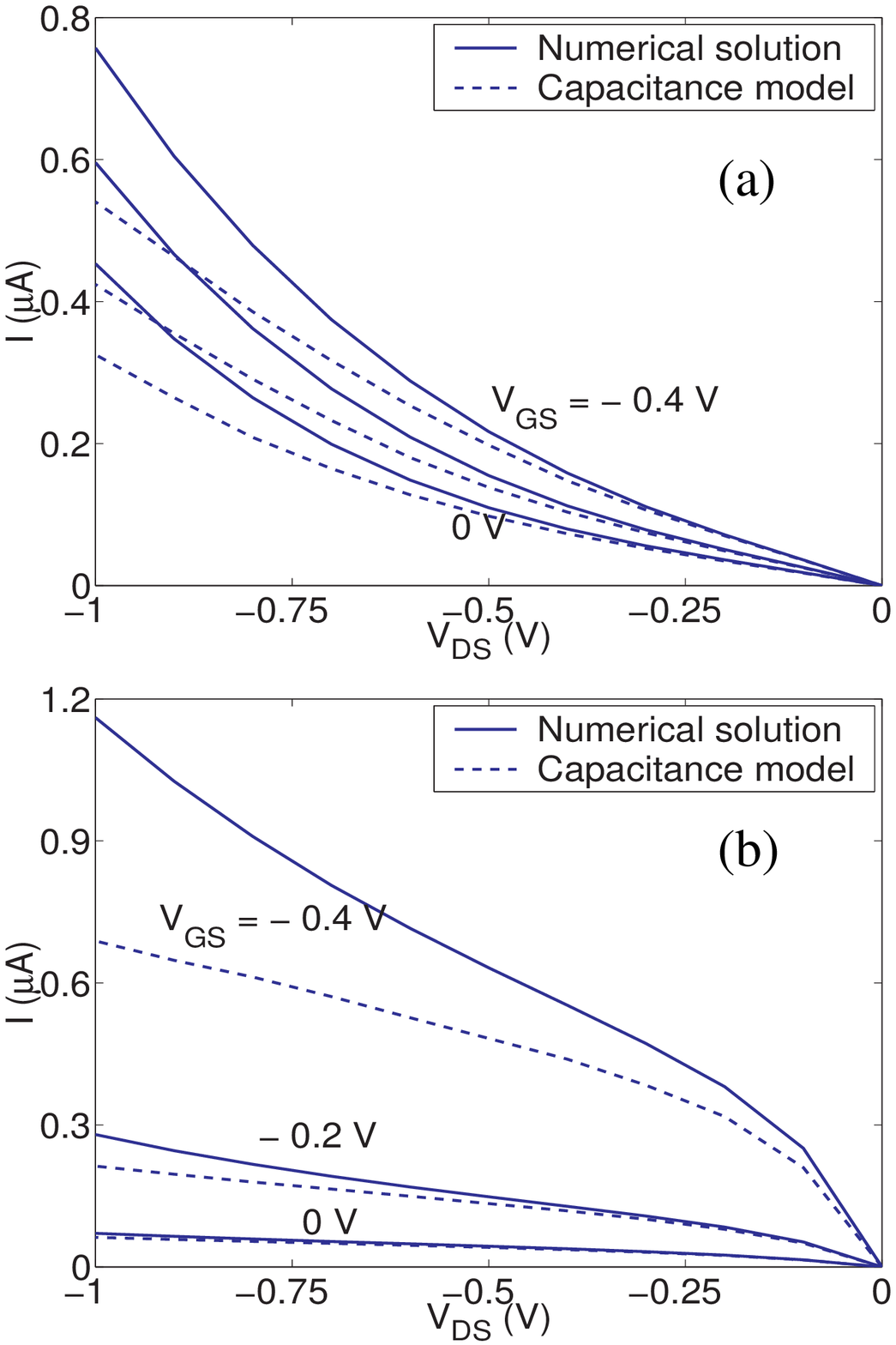,width=3in}}
\caption{
Comparison of the numerical Poisson solution with the capacitance model. We see reasonable agreement in the two solutions
despite the simplifications made in the capacitance model. (a) A realistic case with $t_{ox}=1.5~nm$ which yields $\beta=0.28$
and $U_0=1.9~eV$. No current saturation is seen, but some gate modulation is present. (b) An "absurd" case with $t_{ox}=1~\AA$
which yields $\beta=0.82$ and $U_0=1~eV$ ($U_0$ is less for this case because the gate is closer to the molecule; screening
effect of the gate electrode is larger). The IV for this case looks like that for a MOSFET. Similar IV may be obtained with
$t_{ox}=1~nm$, provided one uses a gate insulator with a dielectric constant about ten times that of silicon dioxide.}
\label{fig:compare}
\end{figure}

Fig.~\ref{fig:compare} compares the IV characteristic obtained by solving the self-consistent NEGF-Poisson's equations with the
numerical Poisson's solution and the capacitance model.  The parameters $\beta$ and $U_0$ for the capacitance model were
extracted from the numerical solution. We see a reasonable agreement between the two solutions despite the simplifications made
in the capacitance model (the capacitance model assumes a flat potential profile in the region occupied by the molecule, which
may not be true, especially at high bias) . For $t_{ox}=1.5~nm$ (Fig.~\ref{fig:compare}a) there is very little gate modulation
and no saturation as expected.  In this case $\beta=0.28$ (Fig.~\ref{fig:beta}) and the IV resembles that shown in
Fig.~\ref{fig:ideal_iv}c more than the one in Fig.~\ref{fig:ideal_iv}a. Also seen in Fig.~\ref{fig:compare} are the results for
$t_{ox}=1~\AA$. For this case $\beta=0.82$ and we observe reasonable saturation and gate control. For realistic oxide
thicknesses, however, we expect to observe an IV like the one shown in Fig.~\ref{fig:compare}a. We have also calculated the IV
characteristics with a double gated geometry (not shown here), and as expected from Fig.~\ref{fig:beta}, saturating IVs can be
obtained for more realistic oxide thicknesses ($\sim ~ 7~\AA$).

\section{Conclusion} 
\label{sec:conclusion}

We have presented simulation results for a three terminal molecular device with a rigid molecule acting as the channel in a
standard MOSFET configuration. The NEGF equations for quantum transport are self-consistently solved with the Poisson's
equation. We conclude the following:

\begin{enumerate}
\item{The current-voltage (IV) characteristics of molecular conductors are strongly influenced by the electrostatics, just like
conventional semiconductors.  With good gate control, the IV characteristics will saturate for one polarity of the drain bias
and increase monotonically if the polarity is reversed. By contrast two-terminal symmetric molecules typically show symmetric IV
characteristics.}
\item{The only advantage gained by using a molecular conductor for an FET channel is due to the reduced dielectric constant of
the molecular environment. To get good gate control with a single gate the gate oxide thickness needs to be less than 10\% of
the channel (molecule) length, whereas in conventional MOSFETs the gate oxide thickness needs to be less than 3\% of the channel
length. With a double gated structure, the respective percentages are 60\% and 20\%.}
\item{Relatively poor subthreshold characteristics (a {\em temperature independent} subthreshold slope much larger than
$60~mV/decade$) are obtained even with good gate control, if metallic contacts (like gold) are used, because the metal-induced
gap states in the channel preclude it from turning off abruptly. Preliminary results with a molecule coupled to doped silicon
source and drain contacts, however, show a temperature dependent subthreshold slope ($\sim k_BT/q$). We believe this is due to
the band-limited nature of the silicon contacts, and we are currently investigating this effect.}
\item{Overall this study suggests that superior saturation and subthreshold characteristics in a molecular FET can only arise
from novel physics beyond that included in our model. Further work on molecular transistors should try to exploit the additional
degrees of freedom afforded by the ``soft'' (as opposed to rigid) nature of molecular conductors \cite{titash}.} \end{enumerate}

{\em Acknowledgments}: We would like to thank M. Samanta, A. Ghosh, R. Venugopal and M. Lundstrom for useful discussions. This
work was supported by the NSF under grants number 9809520-ECS and 0085516-EEC and by the Semiconductor Technology Focus Center
on Materials, Structures and Devices under contract number 1720012625.

%\bibliography{main}

\end{document}